\documentstyle[]{l-aa}

\begin{document}
\input{psfig.sty}

\thesaurus{10.07.2; 11.05.1}

\title{The dichotomy of early--type galaxies from their globular cluster systems 
}

\author {Markus Kissler-Patig \inst{1,2} 
} 

\offprints {M. Kissler-Patig}

\institute{
Sternwarte der Universit\"at Bonn, Auf dem H\"ugel 71, 53121 Bonn, Germany
\and 
European Southern Observatory, Karl-Schwarzschild-Str.~2, 85748 Garching, Germany
}

\date {}

\maketitle

\begin{abstract}
Growing evidence for the existence of two classes of ellipticals calls for a
comparison of the properties of their globular cluster systems.
I compiled information on the properties of 53 globular cluster systems of
faint and bright early-type galaxies, and investigated them in the light
of the properties of the parent galaxies.
The properties of globular cluster systems appear to separate into two 
classes rather than to follow continuous relations with their host galaxy 
properties. 

The ``faint'' systems have a low specific frequency (less than about 5),
less than roughly 1500 globular clusters, a relatively low metallicity
([Fe/H]$ < -1.2$), and a steep density profile that follows the galaxy
light. These systems appear essentially unperturbed, and are hosted 
by faint (about $M_V > -21.5$), disky early--type galaxies with
unresolved cores.

On the other hand, ``bright'' globular cluster systems had a higher
efficiency in producing globular clusters and have higher
specific frequencies (higher than 5). They have larger number of globular
clusters (more than 2000), have flat density profiles ($\alpha > -1.7$)
and their color distributions are often broad, and show several
peaks or gradients in many cases. Finally the mean metallicity is higher
than in ``faint'' globular cluster systems. ``Bright'' globular cluster
systems show all signs predicted for globular cluster systems that
experienced a merger event, and are associated with bright (about $M_V < 
-21.5$) boxy ellipticals with resolved cores. 

I conclude that every galaxy is likely to have formed globular
clusters during the early proto--galactic collapse, but ``bright''
systems were enriched and disturbed during merger events. These two classes 
of globular cluster systems support the idea that major merger events could be 
a cause for the dichotomy of early--type galaxies.


\keywords{globular cluster systems -- globular clusters -- early--type
galaxies }

\end{abstract}

\section{Introduction}

More and more facts support the idea of a dichotomy in early--type
galaxies. Early work from Davies et al.~(1983) already identified kinematically
low luminosity ellipticals ($M_B \ge -20.8$) with the bulges of spiral
galaxies. They distinguished between faint ellipticals that rotate
rapidly and are approximately isotropic and bright ones that rotate
slowly and are anisotropic. Similar distinction came from Nieto (1988) and
Bender (1988), who divided ellipticals in disky and boxy objects (with
possible sub--classes, Nieto \& Bender 1989) from isophotal analyses,
and suggested the division in two classes: disky ellipticals, galaxies
having kept their original structure; and ``merger products'', galaxies
showing signatures of merging of all kinds. This was supported by
the fact that radio--loud galaxies and those surrounded by X--ray halos were 
found to be boxy or irregular, while disky galaxies are mostly radio--quiet
and show no X--ray emission in excess of the discrete source contribution (Bender
et al.~1989). Nieto et al.~(1991) strengthen the similarity between disky
ellipticals and S0 galaxies, having both more peaked, unresolved central 
profiles, while a fair proportion of boxy and irregular ellipticals have
flat, well--resolved central profiles. This idea was confirmed by a
series of papers on ``Centers of Early--Type Galaxies with HST'' in which early--type 
galaxies were found to be divided into two classes based on their
surface--brightness profiles: those with cuspy cores and those whose
steep power--law profiles continue unchanged into the
resolution limit (Lauer et al.~1995, see also
Jaffe et al.~1994 for a similar classification). This bi--modal distribution
was found to correlate with luminosity (Kormendy et al.~1993, Gebhardt et 
al.~1996), defining a bright and faint class (the division occurring at
about $M_V\simeq -21$), which were further
identified with the boxy/non-rotating and disky/rotating classes (Faber
et al.~1996). 

Another approach to this problem is the investigation of the globular
cluster systems of early-type galaxies. Since almost all early--type
galaxies contain a system of old globular clusters (see reviews of Harris 
\& Racine 1979, Harris 1991, Richtler 1995), and globular clusters are believed
to have followed the galaxy formation and evolution, differences in   
formation and evolution of two classes of early--type galaxies might be
visible in the properties of the globular cluster systems. Further 
predictions for the properties of globular cluster systems having
experienced a merger were made by Ashman \& Zepf 1992 to which the observations
can be compared. Finally the properties of a sample of individual objects 
might more clearly show what causes the difference between the two families 
of early--type galaxies. 

Globular cluster systems have been identified in more than 50 early--type
galaxies (see list of Harris \& Harris 1996), from which roughly half
were investigated in more detail. The properties that can be derived are
the number of globular clusters, which can be normalized to the luminosity of
the galaxy (specific frequency $S_N$, see Sect.~3.1), morphological properties
such as surface density profile and spatial distribution, and finally the color
distribution of the globular clusters. 

Various attempts have been made to find correlations between
some of these properties and galaxy parameters. The most extensive study
being the multivariate analysis of Santiago \& Djorgovski (1993) who
looked for relations between number
globular clusters or specific frequency and galaxy parameters,
and concluded that the mechanisms of formation of globular
clusters must be closely tied to those operating during the formation and
evolution of their host galaxies. 
Further studies will be discussed in detail below. 
However the correlations were looked at in sparse samples
available at that time, mainly including bright ellipticals because of
their large globular cluster systems and/or only linear or continuous relations
were assumed.

In Sect.~2 I present the compilation of early--type galaxies with studied 
globular cluster systems. Section 3 shows the correlations between globular
cluster system and galaxy parameters. In Sect.~4 I discuss the results
and their implications.

\section{The compilations of globular cluster systems}

Over 50 early-type galaxies have their globular cluster systems
investigated to date, unfortunately in a very heterogeneous way. 
Table 1 compiles the result for all systems of S0 and ellipticals presently 
known.
The compilation is based on the Harris \& Harris (1996) list, enriched with
information from various sources, especially enclosing four fainter
($M_V \simeq -20.0$) ellipticals from the Fornax cluster 
(Kissler-Patig et al.~1996a), improving the statistics at the faint end.
The reader is referred to Harris \& Harris (1996) and Harris (1991) for 
references when not indicated
differently in column 11 (see below). Differences might occur if articles were
interpreted differently, but I stayed as close as possible to the
original works. 

The following properties were retained for the globular cluster
systems: the total number of globular clusters around the galaxy
($N_{gc}$, column 7) as extrapolated by the authors, with preference for $H_0=75$
km$\cdot$ s$^{-1}\cdot$ Mpc$^{-1}$ when values were given as a function of the
distance; the specific frequency $S_N$ (Number of globular clusters normalized
on the galaxy luminosity, Harris \& van den Bergh 1981) based on the
number of globular clusters and the absolute luminosity given in the
table (column 8); the slope of the density profile ($\alpha$, column 9) as the
exponent of a power--law fitted to the globular cluster density profile;
and finally qualitative color informations (column 10), where ``s''
means single peaked color distribution, ``b'' means broad (or
bi--modal) color distribution (e.g.~Zepf \& Ashman 1993, Kissler-Patig et 
al.~1996a for meaning and implications), and ``g'' means that a color 
gradient was detected in the globular cluster system.
Results marked with a question mark (?) are quoted as uncertain by the
authors. 
The slope of the density profile and color informations were not available for
all galaxies.

For the galaxies, I list the name and type (column 1 and 2); the assumed
distance modulus (column 3) where distances were computed with 
$H_0=75$ km$\cdot$ s$^{-1}\cdot$ Mpc$^{-1}$ only when no other distance
indicator was available; the absolute luminosity (column 4) derived with
the distance stated in column 3; the $a4/a$ coefficient characterizing the 
isophotal shape (column 5) taken from Bender et al.~(1989), Bender et
al.~(1996), and Goudfrooij et al.~(1994), where $a4/a > 0$ indicates
disky isophotes, and $a4/a < 0$ indicates boxy isophotes; and finally 
the core resolution ($\gamma$, column 6) taken from Lauer et al.~(1995).

The galaxies are listed in increasing order of absolute magnitude.	 
The names of the galaxies are marked with a star (*) when less then
10\% of the total number of globular clusters were observed, and thus
the results for these galaxies are rather uncertain.
The references in column 11 refer to: 1,2: reference to be found in the
lists of Harris 1991 and Harris \& Harris 1996 respectively,
3: Ajhar et al.~1994, 4: Forbes et al.~1996 5: Minniti et al.~1996a, 6: 
Kissler-Patig et al.~1996a, 7: Minniti et al.~1996b, 8: Madejski \& Rabolli 1994,
9: Elson \& Santiago 1996, 10: Whitmore et al.~1995, 11: Hilker \& 
Kissler-Patig 1996, 12: Geisler et al.~1996, 13: Bridges et al.~1996,
14: Couture et al.~1991, 15: Dirsch 1996.	
\begin{table*}
\caption{Properties of the globular cluster systems and their parent
galaxies (see Sect.~2)}
\begin{tabular}{|c | c c c c c | c c c c| c|}
\hline
 & \multicolumn{5}{c}{Galaxy parameters}&\multicolumn{4}{c}{GCS
parameters}& \\
 NGC  & T & $(m-M)_V$ & $M_V^T$ & $a4/a$& $\gamma$  & $N_{gc}$ &  $S_N$ & $\alpha$ & {\small Color} & Refs\\
 & & & & (x100) & & & & & & \\
\hline
  artif.& dE    &     0 & -15.0 &      & $               $ & $     6 \pm    0 $& $     6 \pm   0 $& $ -2.2 \pm  0.2$&  & 5 \\
  221   & E2    & 24.54 & -16.3 &      & $               $ & $     3 \pm    0 $& $   0.8 \pm   0 $& $             0$&  & 1 \\
 3115DW1& dE1,N &  30.2 & -17.7 &      & $               $ & $    59 \pm   23 $& $   4.9 \pm 1.9 $& $ -1.8 \pm  0.4$& s  & 2 \\
 3226   & E2    &  31.0 & -19.6 &      & $               $ & $   480 \pm  170 $& $     7 \pm 2.4 $& $ -2.5 \pm  0.5$&  & 1 \\
 4278   & E1    &  30.0 & -19.8 &   -1 & $               $ & $   730 \pm   85 $& $   8.7 \pm 1.4 $& $-1.85 \pm  0.2$& s? & 1,4 \\
 1374   & E1    &  31.0 & -19.8 &    0 & $               $ & $   410 \pm   82 $& $   4.9 \pm 1.3 $& $ -1.8 \pm  0.3$& s  & 6 \\
 1379   & E0    &  31.0 & -19.9 &  0.2 & $               $ & $   314 \pm   63 $& $   3.4 \pm 0.9 $& $ -2.1 \pm  0.6$& s  & 6 \\
 1427   & E3    &  31.0 & -20.0 &  0.7 & $               $ & $   510 \pm   87 $& $   5.1 \pm 1.3 $& $ -2.0  \pm 0.3$& s  & 6 \\
 4340*  & S0    &  31.0 & -20.0 &      & $               $ & $   775 \pm  310 $& $     8 \pm 3.2 $& $              $&  & 1 \\
 4564*  & E6    &  31.0 & -20.1 &  2.2 & $ 1.91 \pm 0.03 $ & $  1000 \pm  300 $& $    10 \pm   3 $& $              $&  & 2 \\
 3377   & E5    &  30.3 & -20.1 &  1.2 & $ 1.95 \pm 0.05 $ & $   235 \pm   50 $& $   2.1 \pm 0.5 $& $ -1.9 \pm  0.2$&  & 1 \\
 1387   & S0    &  31.0 & -20.2 &      & $               $ & $   389 \pm  110 $& $   3.2 \pm 1.1 $& $ -2.2 \pm  0.3$& s  & 6 \\
 5481   & E3    &  32.7 & -20.2 &      & $               $ & $   300 \pm   70 $& $   2.5 \pm 0.6 $& $ -1.7 \pm  0.0$& s? & 8 \\
 3384   & S0    &  30.3 & -20.3 &    0 & $ 2.07 \pm 0.04 $ & $   140 \pm   60 $& $   1.1 \pm 0.5 $& $              $&  & 1 \\
 1052   & E4    &  31.3 & -20.4 &   irr& $               $ & $   430 \pm   80 $& $     3 \pm 0.4 $& $-2.26 \pm 0.27$&  & 1 \\
 3607   & S0    &  30.7 & -20.7 &   irr& $               $ & $   800 \pm  560 $& $   4.2 \pm   3 $& $ -2.6 \pm  0.5$&  & 1 \\
 1549   & E0    &  30.7 & -20.8 & -0.4 & $               $ & $   165 \pm   60 $& $   0.8 \pm 0.3 $& $ -1.8 \pm    0$&  & 1 \\
 5813   & E1    &  31.7 & -21.0 &   irr& $ 0.64 \pm 0.15 $ & $  1800 \pm  400 $& $   7.2 \pm 1.9 $& $-2.18 \pm 0.33$& s  & 2 \\
 3379   & E1    &  30.3 & -21.0 &  0.2 & $ 1.07 \pm 0.06 $ & $   290 \pm  150 $& $   1.2 \pm 0.7 $& $ -1.8 \pm  0.2$&  & 1 \\
 4494   & E0    &  30.8 & -21.0 &  0.3 & $               $ & $  1400 \pm  350 $& $   5.4 \pm 1.3 $& $-1.06 \pm 0.38$& b? & 2,4 \\
 1404   & E1    &  31.0 & -21.0 &  0.5 & $               $ & $   880 \pm  140 $& $   3.5 \pm 0.8 $& $   -2 \pm  0.1$& s  & 2 \\
 1553   & S0    &  30.5 & -21.0 &      & $               $ & $   587 \pm  130 $& $   2.3 \pm 0.5 $& $ -2.3 \pm    0$&  & 1 \\
 3115   & S0    &  30.2 & -21.1 &      & $ 1.87 \pm 0.03 $ & $   630 \pm  150 $& $   2.3 \pm 0.5 $& $-1.84 \pm 0.23$&  & 1 \\
  720   & E5    &  31.4 & -21.2 &  0.7 & $ 0.73 \pm 0.32 $ & $   660 \pm  190 $& $   2.2 \pm 0.9 $& $ -2.2 \pm  0.1$&  & 2 \\
 4552*  & E0    &  31.0 & -21.2 &   -2 & $ 0.63 \pm 0.14 $ & $  2400 \pm    0 $& $     8 \pm   0 $& $              $& b? & 2,3 \\
 4621*  & E5    &  31.0 & -21.2 &  1.5 & $ 2.03 \pm 0.03 $ & $  1900 \pm  400 $& $   6.3 \pm 1.2 $& $              $&  & 1 \\
 4526*  & S0    &  31.0 & -21.4 &      & $               $ & $  2700 \pm  400 $& $   7.7 \pm 1.2 $& $              $&  & 1 \\
 4697   & E6    &  30.8 & -21.6 &  1.4 & $               $ & $  1090 \pm  420 $& $   2.5 \pm 1.0 $& $ -1.9 \pm 0.2$ &  & 15 \\
 4881   & E0    &  34.9 & -21.6 &      & $               $ & $   390 \pm   40 $& $     1 \pm 0.1 $& $             0$&  & 2 \\
 4636   & E0    &  31.2 & -21.7 & -0.2 & $    1 \pm 0.09 $ & $  3600 \pm  500 $& $   7.5 \pm   2 $& $   -1 \pm  0.1$&  & 2 \\
 4374*  & E1    &  31.0 & -21.7 & -0.4 & $               $ & $  3040 \pm  400 $& $   6.6 \pm 0.9 $& $              $& s? & 1,3 \\
 1399   & E1/cD &  31.0 & -21.7 &  0.1 & $ 0.78 \pm 0.14 $ & $  5940 \pm  570 $& $  12.4 \pm   3 $& $ -1.6 \pm 0.15$& bg & 2,6 \\
 5629   & E/cD  &  34.0 & -21.7 &      & $               $ & $  2000 \pm    0 $& $     5 \pm   0 $& $              $&  & 2 \\
 4406*  & E3    &  31.0 & -21.8 & -0.7 & $               $ & $  3350 \pm  400 $& $   6.3 \pm 0.8 $& $              $& s? & 1,3 \\
 4365   & E2    &  31.4 & -21.8 & -1.1 & $ 1.06 \pm 0.09 $ & $  2500 \pm  200 $& $     5 \pm 0.4 $& $-1.15 \pm 0.25$& bg & 2,3 \\
  524   & S0    &  32.5 & -21.9 &      & $               $ & $  3300 \pm 1000 $& $   4.8 \pm 1.1 $& $-1.71 \pm 0.12$&  & 1 \\
 5128   & E0p   & 28.25 & -22.0 &      & $               $ & $  1700 \pm  400 $& $   2.6 \pm 0.6 $& $ -1.5 \pm  0.2$& bg & 1,7 \\
 5846   & E0    &  32.3 & -22.1 &  0.0 & $               $ & $  3120 \pm 1850 $& $   4.5 \pm 2.7 $& $              $&  & 1 \\
 3923   & E3    &  31.9 & -22.1 & -0.4 & $               $ & $  4300 \pm 1000 $& $   6.4 \pm 1.5 $& $              $& bg & 2 \\
 4649*  & E2    &  31.0 & -22.2 & -0.5 & $ 0.82 \pm 0.16 $ & $  5100 \pm  160 $& $   6.9 \pm 0.2 $& $              $&  g & 1,14 \\
 3311   & E0/cD &  33.4 & -22.3 &      & $               $ & $ 12400 \pm 5000 $& $    15 \pm   6 $& $ -1.3 \pm  0.2$& b  & 2 \\
 4486   & E0    &  31.0 & -22.4 &    0 & $               $ & $ 13000 \pm  500 $& $  13.9 \pm 0.5 $& $-1.61 \pm 0.08$& bg & 2,9,10 \\
 5018   & E4p   &  33.4 & -22.6 &      & $               $ & $  1200 \pm  500 $& $   1.1 \pm 0.5 $& $ -1.3 \pm  0.4$& b  & 11 \\
 3557*  & E3    &  33.0 & -22.6 &   0.0& $               $ & $   400 \pm  300 $& $   0.4 \pm 0.3 $& $              $&  & 1 \\
 4472   & E2    &  31.0 & -22.6 & -0.3 & $  0.9 \pm 0.19 $ & $  6300 \pm 1900 $& $   5.6 \pm 1.7 $& $-1.68 \pm 0.13$& bg & 1,3,12 \\
 6166   & E2/cD &  35.45& -22.74&      & $               $ & $ 11000 \pm 8000 $& $     9 \pm   6 $& $ -0.95\pm  0.1$&s?g?& 2,13\\
 7768   & E2/cD &  35.3 & -22.9 &      & $               $ & $  4050 \pm 2600 $& $   2.8 \pm 1.8 $& $ -1.3 \pm    0$&  & 2 \\
 4874   & E0    &  34.9 & -23.0 &      & $ 0.76 \pm 0.17 $ & $ 22600 \pm 2700 $& $  14.3 \pm 1.7 $& $              $&  & 2 \\
 4073   & E1/cD &  34.6 & -23.1 &      & $               $ & $  8290 \pm  460 $& $   4.8 \pm 0.3 $& $-0.95 \pm  0.3$&  & 2 \\
 3842   & E3    &  34.7 & -23.1 & -0.3 & $               $ & $ 14000 \pm 2500 $& $   7.7 \pm 1.4 $& $ -1.2 \pm    0$&  & 2 \\
 1275   & Ep/cD &  34.9 & -23.3 &      & $               $ & $  7750 \pm 2520 $& $   4.3 \pm 1.4 $& $              $& b  & 2 \\
UGC9958 & E/cD  &  36.3 & -23.4 &      & $               $ & $ 27000 \pm13000 $& $    12 \pm 5.6 $& $ -1.2 \pm    0$&  & 2 \\
UGC9799 & E/cD  &  35.9 & -23.4 &      & $               $ & $ 48000 \pm16000 $& $    21 \pm   7 $& $ -1.4 \pm    0$&  & 2 \\
 4889   & E4    &  34.9 & -23.5 &   irr& $ 0.33 \pm 0.44 $ & $ 17300 \pm 3000 $& $   6.9 \pm 1.2 $& $              $&  & 2 \\
\hline
\end{tabular}
\end{table*}

\section{Globular cluster system vs.~galaxy parameters}

\subsection{The specific frequency}

The specific frequency $S_N=N\times10^{(0.4\cdot(M_V + 15))}$, where $N$
is the total number of globular clusters, and $M_V$ the absolute
luminosity of the galaxy, was introduced by Harris \& van
den Bergh (1981). It was thereafter often used to look for correlations of 
globular cluster system properties with galaxy properties (e.g.~the extensive 
analysis by Santiago \& Djorgovski 1993), mainly being interpreted as
a possible influence of the galaxy environment on the efficiency of forming
globular cluster (e.g.~Harris 1991, Kumai et al.~1993 for a quantitative
relation). The outstanding ($>10$) specific frequencies of some very
luminous central ellipticals still wait for a definitive explanation,
but were shown to be compatible with a scenario were these galaxies were
formed through major merger events (e.g.~Zepf \& Ashman 1993).

We show the correlations of $S_N$ versus the total luminosity of the galaxy 
$M_V$, and the $a4/a$ coefficient of the isophotal analysis of the
galaxy in Fig.~1 (upper and lower panel respectively). Values from galaxies
marked with a star in Table 1
(i.e.~with less then 10\% of the total number of globular clusters
observed) are plotted as open circles, the others as dots. 
\begin{figure}
\psfig{figure=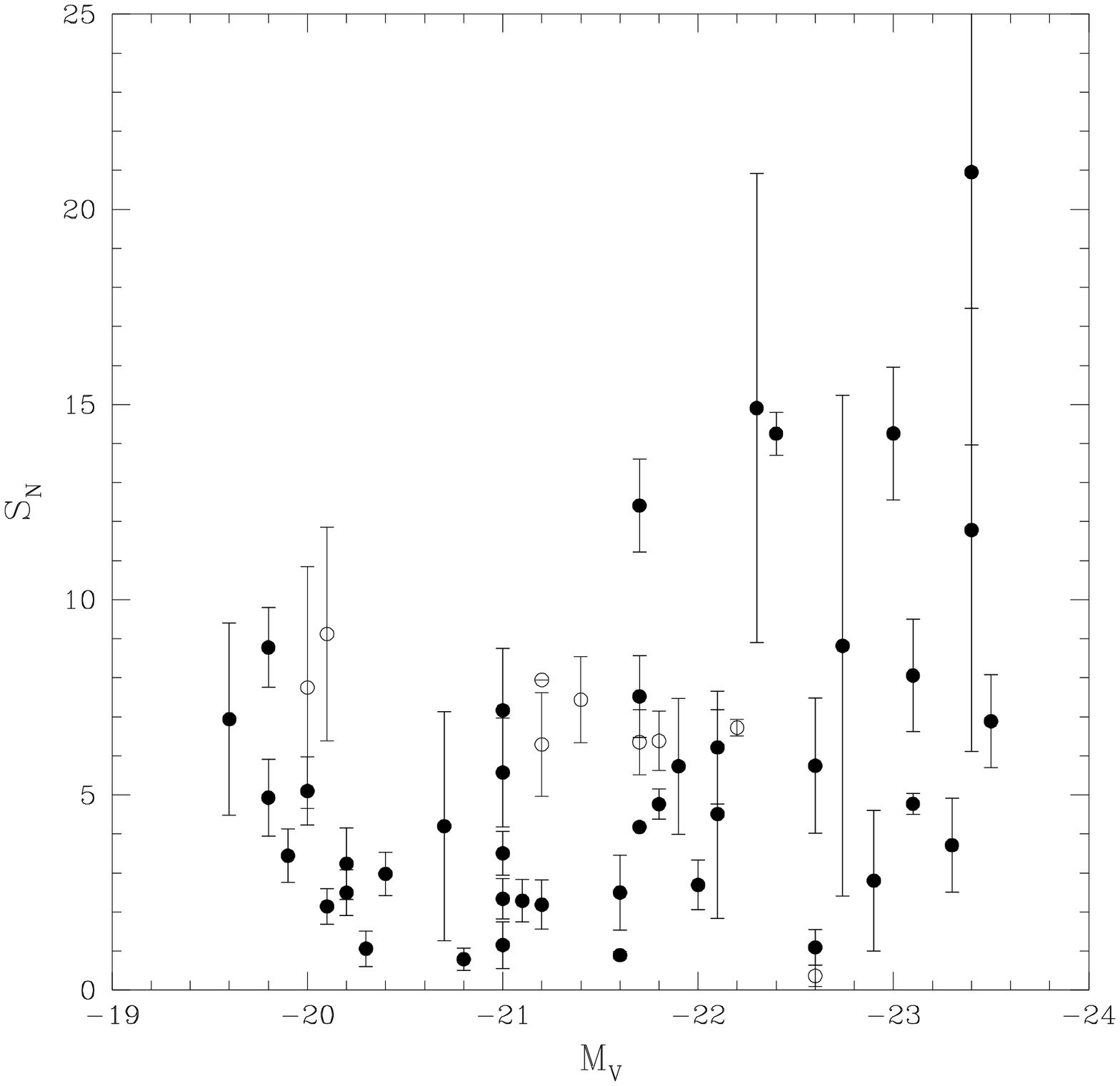,height=8cm,width=8cm
,bbllx=8mm,bblly=57mm,bburx=205mm,bbury=245mm}
\psfig{figure=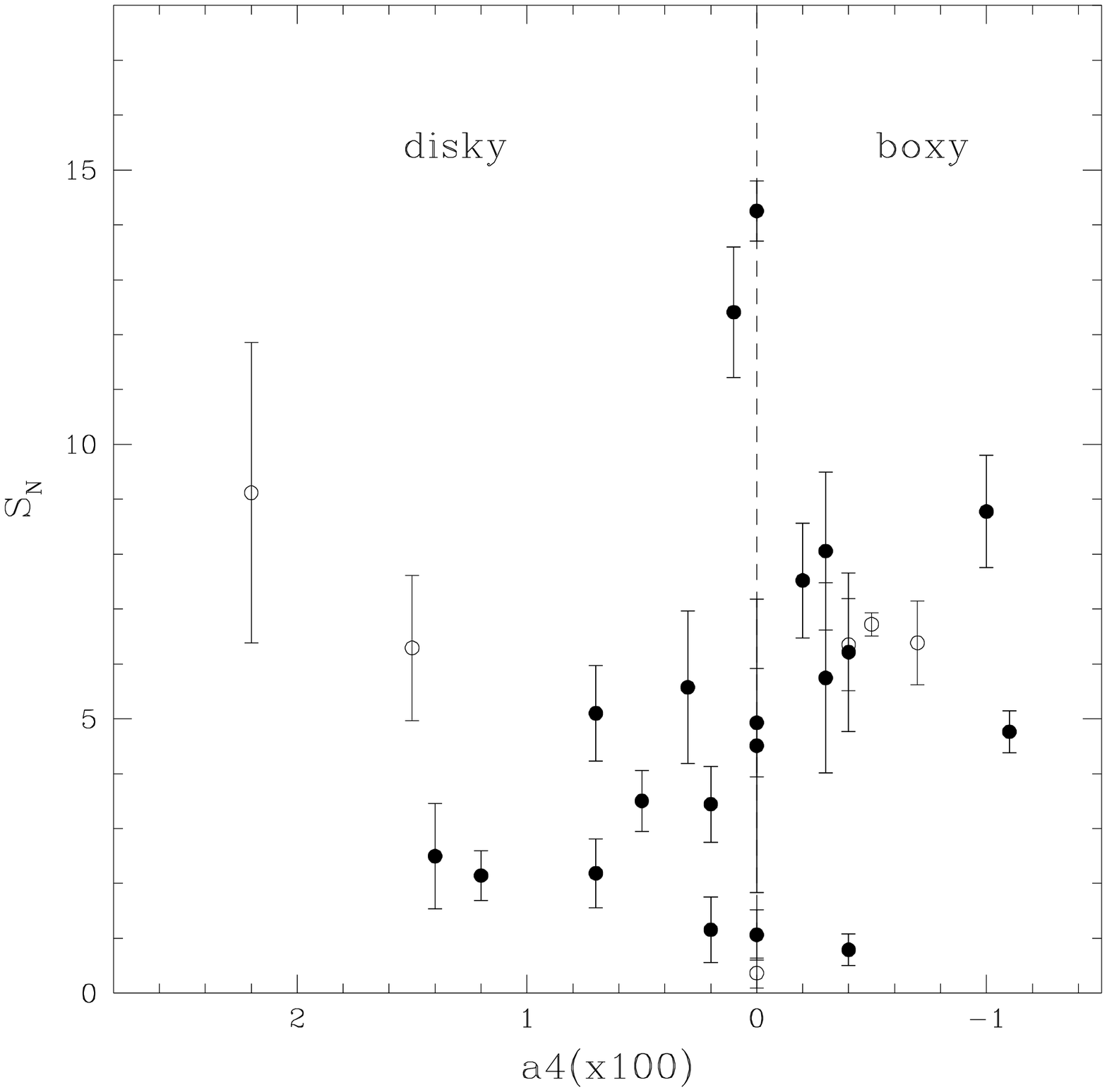,height=8cm,width=8cm
,bbllx=8mm,bblly=57mm,bburx=205mm,bbury=245mm}
\caption{Upper panel: The specific frequency $S_N$ plotted versus the absolute
luminosity of the parent galaxy $M_V$. 
Lower panel: $S_N$ versus the $a4/a\cdot100$ coefficient of the isophotal analysis
of the galaxy ($a4/a > 0$ indicates disky isophotes, $a4/a < 0$ indicates
boxy isophotes). Open circles mark the globular
cluster systems for which less than 10\% of the total number of globular
clusters was observed.
}
\end {figure}

Fig.~1 upper panel, $S_N$ vs $M_V$, does not show any clear correlation.
However note that $S_N$ spans a narrower range for 
faint galaxies; 17 out of 21 galaxies with $M_V > -21.6$ mag and well 
determined total numbers of globular clusters have $S_N$ values of 5 and
below with a mean of 4.0. For more luminous galaxies the mean specific 
frequency is much higher (at 7.5) and
only 4 out of 21 galaxies with $M_V < -21.6$ have $S_N$ values below
4. Further the dispersion in $S_N$ is twice as large for bright 
galaxies than for faint ones.

We tested whether the specific frequency rather follows a continuous
relation with luminosity than representing two classes of ellipticals
with different globular cluster formation efficiencies.
The division between the two classes of ellipticals galaxies occurs between 
$M_V=-21$ and $-22$ mag (see references in introduction). We arbitrarely 
divided our sample at $M_V\simeq-21.5$
(the result is not significantly influenced by small shifts of this
limit) and tested the fit by 
a linear relation between $S_N$ and $M_V$ over the full luminosity range 
against a composite function with a constant $S_N=4$ down to $M_V\simeq-21.5$ and a
linear fit at higher luminosities. While a F--test slightly prefers the
composite model, Kolmogorov-Smirnov (hereafter K--S) tests rather rejects (80\%
confidence) linear relations over the bright or full range. Only the
faint end seems to be marginally well represented by a constant value.

The correlation with the $a4/a$ coefficient (Fig.~1 lower panel) is clearer:
disky galaxies have small $S_N$ values (around 5 and below), while boxy
galaxies have high $S_N$ values (at least 5 and up to 10). This
partly reflects the fact that disky ellipticals are generally fainter
than boxy ellipticals (Bender et al.~1989). Environmental
effects should not significantly influence the result since an equal
amount of disky and boxy galaxies here are galaxy cluster members.
The peculiar $S_N$ values of central giant ellipticals are  
apparent in our plot with NGC 1399 (central gE in Fornax, $S_N\simeq12$) and 
M87 (gE in the center of Virgo, $S_N\simeq14$) that clearly drop out of
the sample.  A K--S test rejects with 95\% confidence the hypothesis that disky 
and boxy ellipticals have similar specific frequencies.  

In summary: if the specific frequency is interpreted as an efficiency of 
producing globular clusters, bright, boxy ellipticals must have been more 
efficient in forming globular clusters than faint, disky ellipticals, or must 
have been enriched by some process, in addition to any possible environmental 
effect. 

\subsection{The number of globular clusters}

The specific frequency relates the number of globular clusters to the
absolute luminosity of the parent galaxy. One can check if the apparent
relations of $S_N$ with galaxy parameters still hold for absolute
numbers of globular clusters. Zepf et al.~(1993) recently discussed a
possible increase of the number of globular cluster ($N_{gc}$) with increasing 
galaxy luminosity and agreed with Djorgovski \& Santiago (1992) that $N_{gc}$
increases with $L$ to a power greater than one. Zepf et al.~(1993) interprete
the trend as the increase of importance of dissipationless merging with
increasing luminosities in the frame of gE being products of mergers.

In figure 2 we show $N_{gc}$, the total number of globular clusters 
versus $M_V$ and $a4/a$.
\begin{figure}
\psfig{figure=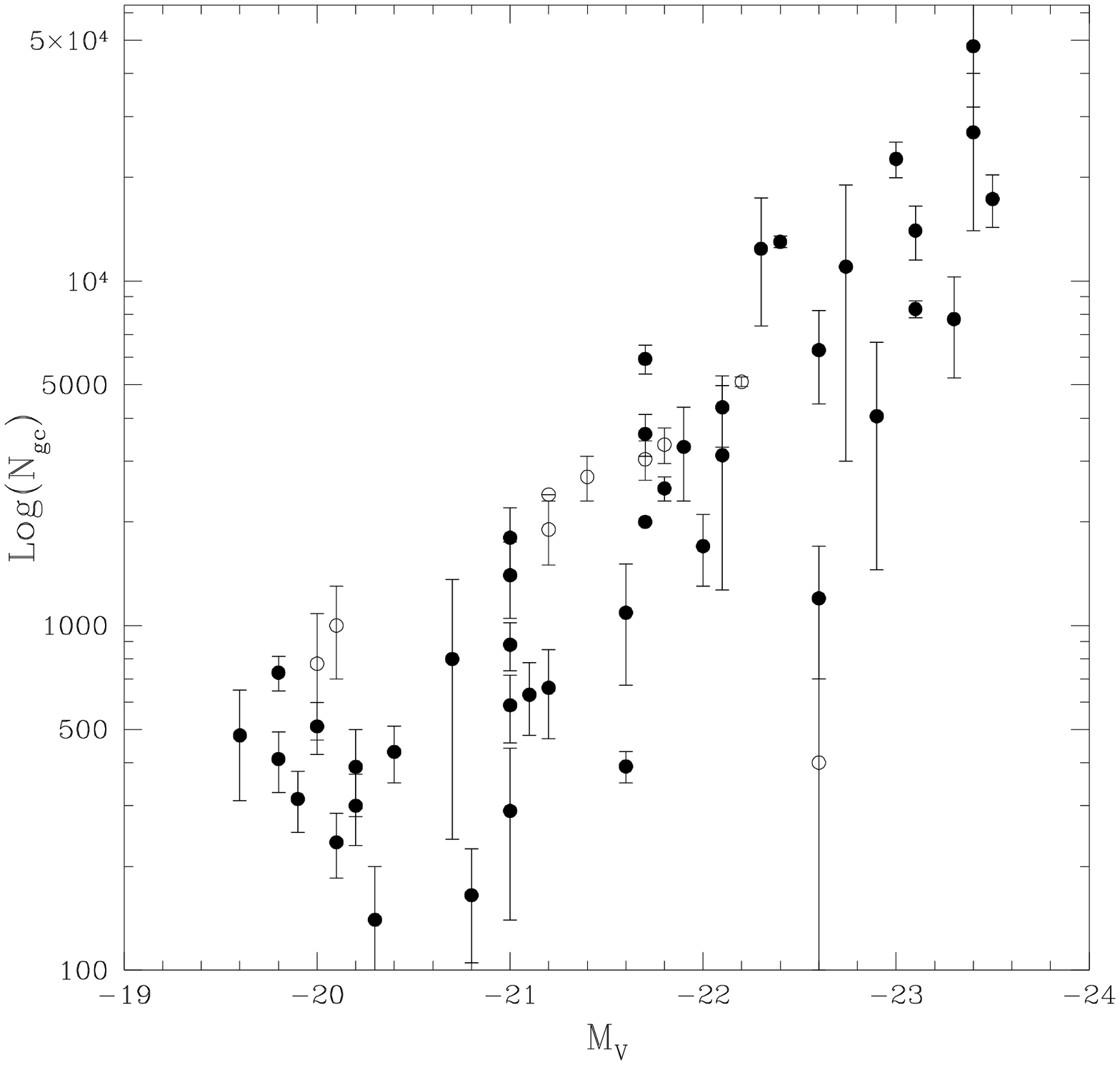,height=8cm,width=8cm
,bbllx=8mm,bblly=57mm,bburx=205mm,bbury=245mm}
\psfig{figure=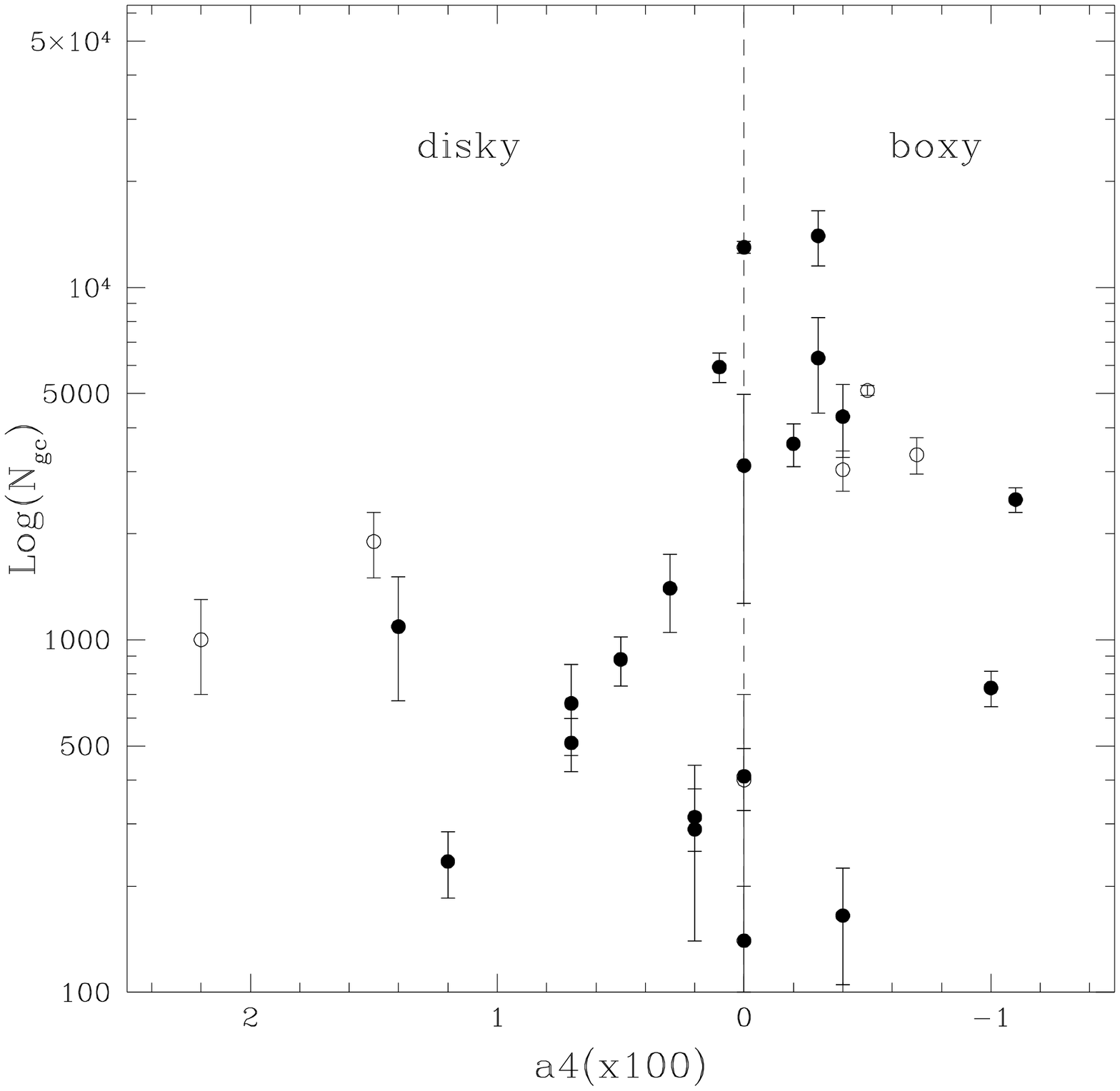,height=8cm,width=8cm
,bbllx=8mm,bblly=57mm,bburx=205mm,bbury=245mm}
\caption{Upper panel: Total number of globular clusters $N_{gc}$ versus absolute
magnitude of the galaxy.
Lower panel: Total number of globular clusters $N_{gc}$ versus the $a4/a$
coefficient from the isophotal analysis of the galaxy ($a4/a > 0$
indicates disky isophotes, $a4/a < 0$ indicates boxy isophotes). 
Open circles mark the globular cluster systems for which less than 10\% of 
the total number of globular clusters was observed.
}
\end {figure}
One is tempted to fit a linear relation in Fig.~2 upper panel. The resulting fit
returns a slope of $-0.50\pm0.05$, which physically is the exponent of a 
power--law since two logarithmic values are plotted versus each other. 
Keeping in mind that the sample might divide around $M_V\simeq -21.5$ (see 
introduction and last section), one could fit linear
relations to the two samples (fainter and brighter than $M_V\simeq-21.5$). The
resulting slopes are flatter ($-0.13\pm0.06$) for the faint galaxies and 
slightly steeper ($-0.56\pm0.05$) for the bright galaxies. Here again a
F--test slightly prefers (60\% confidence) the composite model to a
single law over all magnitudes. Further, a K--S test clearly favors the
composite model as a representation of the data (68\% confidence) to a
continuous relation that is only a good representation with 49\%
confidence. 

Interestingly the composite relation is then no longer
continuous around $M_V=-21.5$ but the zeropoint jumps up by $500\pm200$ 
clusters between the two samples when going to brighter galaxies.

A non--continuous relation is also supported by the lower panel
in Fig.~2 ($N_{gc}$ versus $a4/a$) that, similar to Fig.~1 lower panel, 
clearly divides the systems: while all disky galaxies with well
determined total numbers of globular clusters have
less than 1500 clusters, all boxy galaxies but two have more than 2500
clusters.

In summary, faint, disky ellipticals host less globular clusters than
bright, boxy ellipticals do, and two classes of ellipticals with a
likely jump in their number of globular clusters between the two are 
preferred to a continuous relation with luminosity. Note however that
Djorgovski \& Santiago's (1992) and Zepf's et al.~(1993) interpretation 
still holds for the bright ellipticals. 

\subsection{The slope of the surface density profile}

The question rises whether the number of globular clusters (or the
related $S_N$) is the only property by which globular cluster systems
in bright, boxy and faint, disky early--type galaxies might divide.
The slope of the globular cluster surface density profile $\alpha$ was
suspected to be related to the galaxy luminosity by Harris (1986, 1993)
who used sparser sample available to that time. His linear relation
for globular cluster systems seemed to be displaced to lower central
concentrations with respect to the galaxy light profiles. Harris 
proposed it to trace the formation of globular cluster systems in the
still--extended proto--galaxies. Since then, however, especially faint galaxies
were shown to have globular cluster systems with profiles (as well as 
ellipticities and position angles) compatible with the galaxy light
(e.g.~Kissler--Patig et al.~1996b and references therein).

The correlations of $\alpha$ with the galaxy properties ($M_V$ and
$a4/a$) are shown in Fig.~3.
\begin{figure}
\psfig{figure=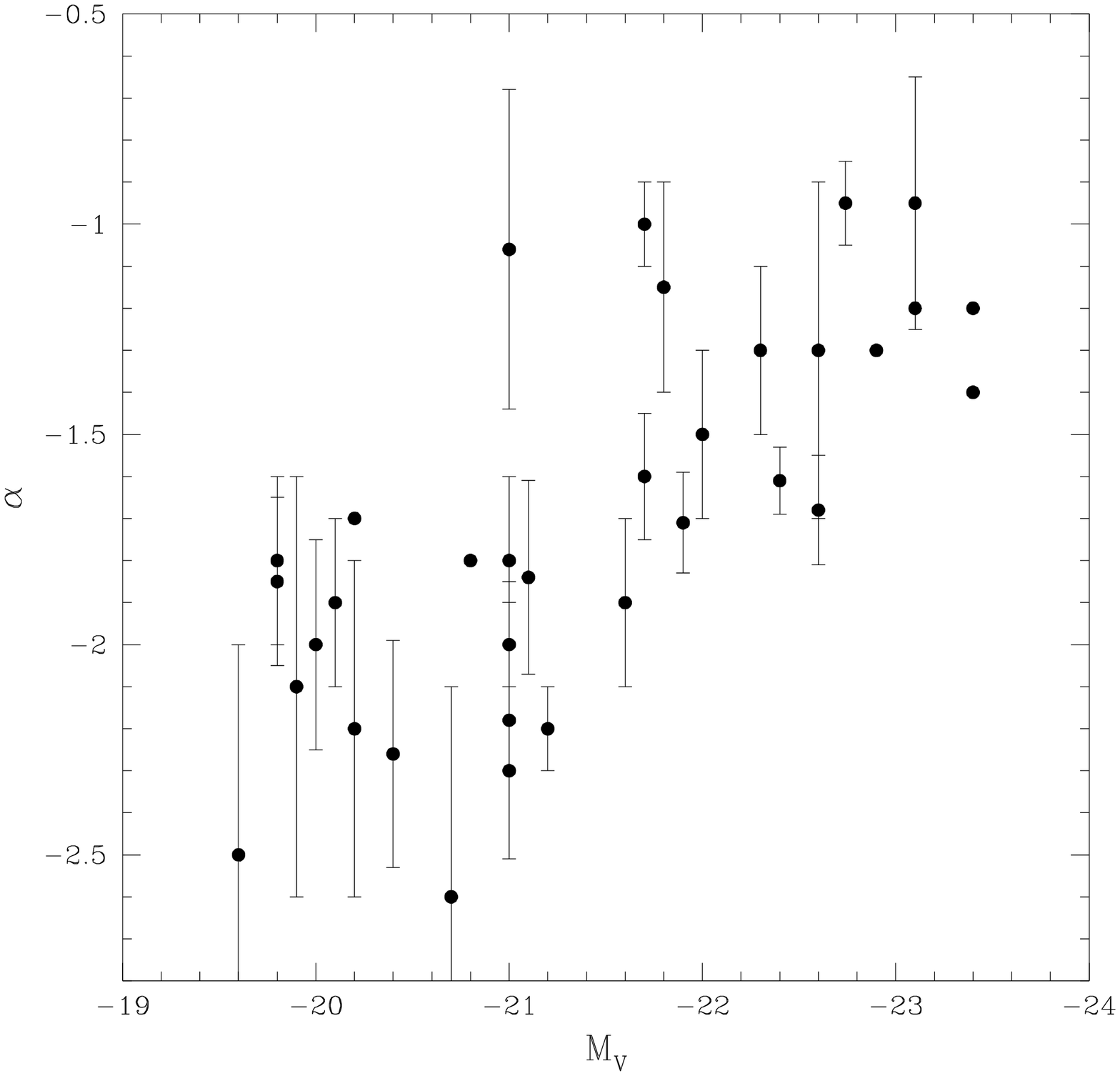,height=8cm,width=8cm
,bbllx=8mm,bblly=57mm,bburx=205mm,bbury=245mm}
\psfig{figure=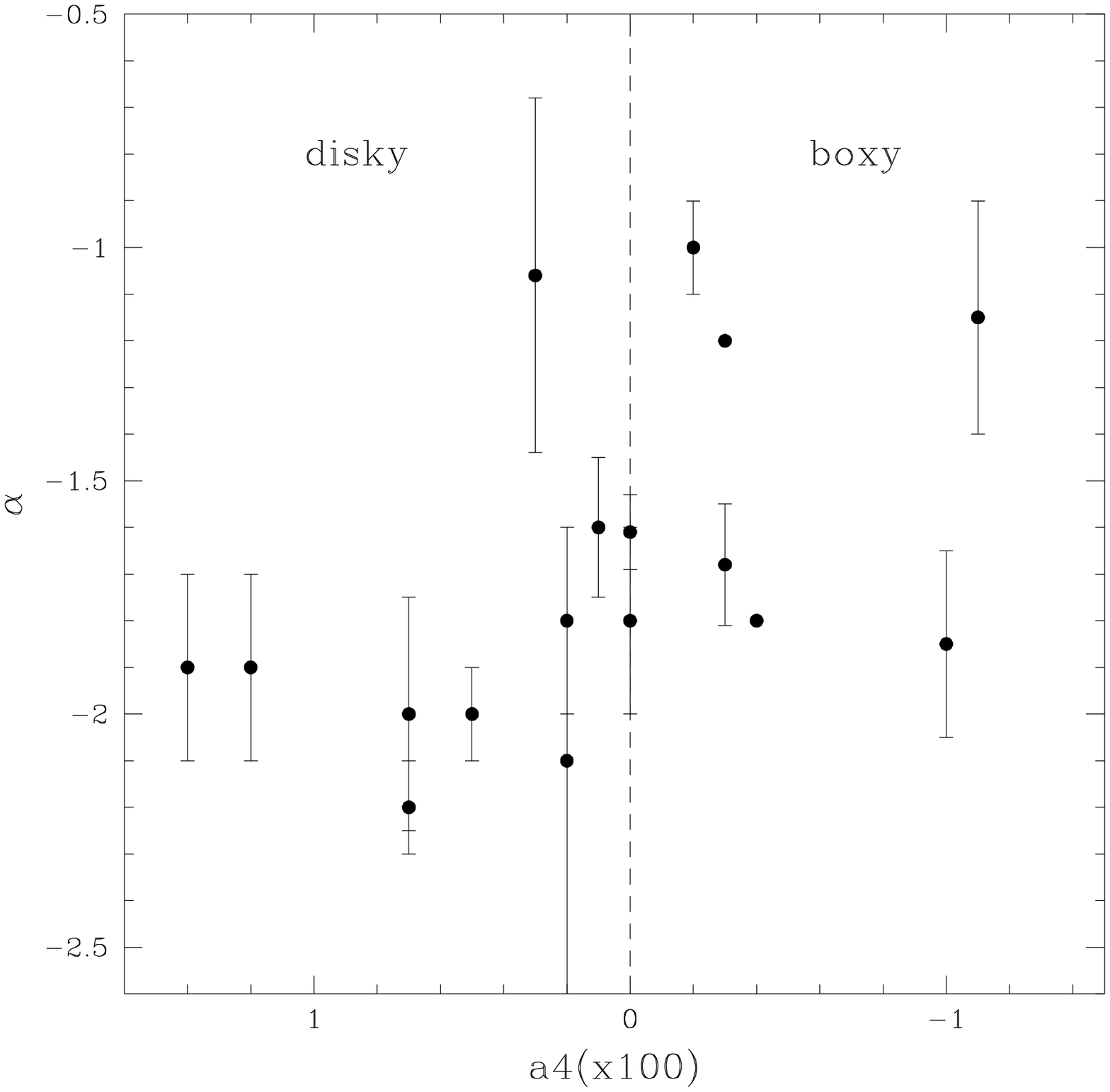,height=8cm,width=8cm
,bbllx=8mm,bblly=57mm,bburx=205mm,bbury=245mm}
\caption{Upper panel: The slope of the surface density profile of the globular
cluster system ($\alpha$) plotted against the absolute luminosity of the
parent galaxy ($M_V$).
Lower panel: The slope of the surface density profile of the globular
cluster system ($\alpha$) plotted against the $a4/a$ value of the parent
galaxy ($a4/a > 0$ indicates disky isophotes, $a4/a < 0$ indicates
boxy isophotes).
}
\end {figure}
The slope of the surface density profile $\alpha$ turns out to be a strong 
discriminator between faint and bright, and disky and boxy ellipticals.
Both upper and lower panel in Fig.~3 are roughly cut into four quadrants from 
which only two are populated: globular cluster systems with slope $\alpha
< -1.8$ only appear in faint ($M_V > -21.5$) and disky
ellipticals, while globular cluster systems in bright, boxy ellipticals
have systematically flatter surface density profiles. K--S test rejects
with more than 95\% and 90\% confidence respectively faint and bright, and
disky and boxy ellipticals to have similar $\alpha$ values. Further an
F--test prefers (70\% confidence) two populations of ellipticals
concentrated around mean $\alpha$ values of $-2.0\pm0.3$ and $-1.3\pm0.3$ 
rather than a linear relation with luminosity.

Faint, disky ellipticals seem to have globular cluster systems
compatible with their light profiles, while bright, boxy ellipticals
have more extended systems.

In figure 4 we plot a consequence of this relation: the total number of 
globular clusters $N_{gc}$ versus the slope of the surface density 
profile $\alpha$. 
\begin{figure}
\psfig{figure=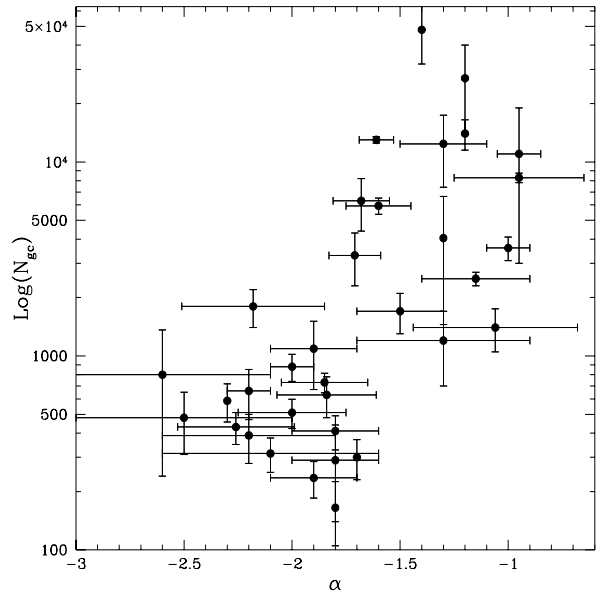,height=8cm,width=8cm
,bbllx=8mm,bblly=57mm,bburx=205mm,bbury=245mm}
\caption{The total number of globular clusters $N_{gc}$ plotted against
the slope of the density profile $\alpha$ of the globular cluster
systems. 
}
\end {figure}
The diagram shows that systems where the globular clusters have
steep ($\alpha < -1.8$) density profiles and follow the galaxy light
correspond to the systems with few ($N_{gc} < 1500$) globular clusters,
while systems with flat ($\alpha > -1.8$) density profiles correspond to richer systems.

\subsection{More globular cluster parameters}

\subsubsection{The mean metallicity}

Perelmuter (1995), Forbes et al.~(1996), and Bridges et al.~(1996) recently 
recompiled metallicities of globular cluster systems and plotted them against 
the mass or absolute luminosity of the parent galaxy (their Figs.~2, 
14, and 7 respectively). Perelmuter (1995) sees a discrepancy between
spirals and ellipticals in the [Fe/H]--mass relation, but used only
bright ellipticals in our sense for his relation. Forbes et al.~(1996) see a
correlation over 10 orders of magnitudes and take it as a strong argument
for the formation of globular clusters during an early collapse. 
The plot of Bridges et al.~(1996), showing more data
towards bright ellipticals, rather implies a smooth rise of
the metallicity from [Fe/H]$\simeq -2$ to $-1.2$ for galaxies from $M_V\simeq
-14$ to $M_V\simeq -21$ mag, followed by a steep up--turn and larger
dispersion over the brightest magnitudes with metallicity values up to near 
solar metallicities.

Thus metallicity also seems to be a parameter compatible with the division 
of the globular cluster systems in the sense that systems in faint
galaxies have low ([Fe/H]$ < -1.2$) mean metallicities correlated with
the galaxy luminosity, while systems in
bright ellipticals have higher mean metallicities and a larger
dispersion.

\subsubsection{The core radius of the globular cluster system}

Forbes et al.~(1996) determined the core radius of the globular
cluster system (introduced by Lauer \& Kormendy 1986) for their 14
ellipticals. They plotted them in their Fig.~10 versus the absolute
magnitude of the galaxy. 
The authors claim only a weak correlation with luminosity and propose that
most luminous ($M_V < -21$) galaxies have large core radii while 
the core radius of systems in small, rotating ellipticals is weakly, if
at all, dependent on galaxy luminosity.
This relation might be expected after Sect.~3.3, since core radius
and surface density profile slope both describe the radial
concentration of the globular clusters in the galaxy, and are
therefore correlated. From a practical point of view, determining core
radii involves observations towards the center and therefore requires very good
seeing or HST observations, while surface density profiles can be
estimated from normal ground--based observations.

\section{The dichotomy of globular cluster systems}

\subsection{Two classes of globular cluster systems}

Two points were shown in section 3. First, two classes of globular cluster 
systems are preferred to continuous relations when their 
properties are investigated versus the absolute luminosity of the parent
early--type galaxies. The ``cut'' systematically appears at about $M_V\simeq
-21.5$ mag. In bright ellipticals the specific frequencies and mean
metallicity of globular clusters show large dispersions up to extremely
high values. Further, the number of globular clusters seem to slightly 
``jump up'' between faint and bright galaxies, and while in faint
galaxies the globular cluster systems have surface density profiles
compatible with the galaxy light, they appear flatter in bright
galaxies.

Second, the globular cluster systems systematically split into two groups 
when their properties are investigated in disky and boxy ellipticals. In disky
galaxies globular cluster systems have lower specific frequencies, less
globular clusters, and steeper density profiles than in boxy ellipticals. 

We can further consider the information on the color distribution
of the globular system (column 10 in Table 1) which shows that broad 
distributions and color gradients appear almost exclusively in bright 
ellipticals.

The above points suggest that globular cluster systems should rather be divided 
into two classes than assumed to form one group with properties scaling with 
the properties of their host galaxies.

The properties of the ``faint'' systems are: a small number of globular
clusters (less than 1500), the related small specific frequency (less
than 5, mean of our sample 2 to 3), a low mean metallicity of the globular 
clusters ([Fe/H]$ <
-1.2$), and a steep surface density profile of the globular clusters
that follows the surface intensity profile of the host galaxy. 

The properties of ``bright'' globular cluster systems are: a high number
of globular clusters (more than 2500), the unproportionaly high specific
frequency (from 5 to 15), a higher mean metallicity than would imply the
relation up to $M_V\simeq -21$, a flat surface density profile ($\alpha
> -1.7$) only compatible with the galaxy light when the galaxy has a cD
envelope, and in most cases broad color distributions or color gradients.

These two different types of systems can be associated with the two
classes of ellipticals that also emerged from recent studies of large
sample of elliptical galaxies.

The ``faint'' globular cluster systems are hosted by faint (about $M_V > -21.5$
mag), disky galaxies with unresolved cores.

The ``bright'' globular cluster systems are hosted by bright (about $M_V
< -21.5$ mag), boxy galaxies with resolved cores.

\subsection{The interpretation}

Various alternatives for the formation of globular clusters and their
over--abundance in some galaxies were proposed. Some seem already rejected
in the case of whole globular cluster systems,
as the pre--galactic formation (Peebles \& Dicke 1968) since no
correlation with the later galaxies is then expected but seen. Some are 
believed to have played only a minor role, such as globular cluster stripping
(Muzzio et al.~1984, Muzzio 1986), and the formation of globular
clusters in cooling flows (e.g.~discussion in Bridges et al.~1996), that
appear to be able to contribute only few globular clusters.

The most popular remaining scenarios are the formation of globular
clusters during the collapse of the proto--galaxy (e.g.~Searle \& Zinn 1978, 
Harris \& Pudritz 1994), and the formation during mergers
(Burstein 1987, Schweizer 1987, Ashman \& Zepf 1992). Two classes of 
ellipticals do not argue for nor
against one of these scenarios, it allows them to co--exist,
since most properties that support the one or other formation
mechanism dominates in one group. 

\subsubsection {``Faint'' systems}

Forbes et al.~(1996) argue that the correlation of the mean metallicity of
the globular clusters with the luminosity of the parent galaxy is a strong
argument for the formation during the early collapse of the galaxy and
the coalescent phase at high redshift. This relation however seems
weakened towards brighter galaxies.

Further the recent investigation globular clusters of faint ($-19<M_V<-21$ mag)
early--type galaxies allowed a better statistic of ``normal'' properties of
globular cluster systems. ``Faint'' systems were shown to
spatially follow the galaxy light rather than being more extended (see
Sect.~3.3) and suppressed the apparent discrepancy between globular cluster and
star formation epoch in these systems. The properties of ``faint''
systems do no call for any formation mechanism beside the formation in
the early collapse of the galaxy (Searle \& Zinn  1978). Faint galaxies appear
unperturbed from the properties of their globular cluster systems, as
well as from dynamical point of view (see references in Sect.~1).

\subsubsection{``Bright'' systems}

The predictions of Ashman \& Zepf (1992) for globular clusters systems
that formed in mergers are the following: New globular clusters should
form; the newly formed clusters should be more metal--rich, since formed from 
processed material;
the system should show a color gradient due to the new (metal--richer)
globular clusters concentrating to the center; the color distribution
should become multi--modal for the same reasons; and finally the surface
density profile should be flatter due to the fact that star formation
occurs in the center concentrating the light distribution, leaving the
old globular clusters (if dominant) appear flatter than the light
profile.

The formation of globular clusters during mergers is uncontested since several 
galaxies were found to form globular clusters in a merger event
(NGC 3597: Lutz 1991; NGC 1275: Holtzman et al. ~1992; NGC
7252: Whitmore et al.~1993, Schweizer \& Seitzer 1993; He2--10: Conti \&
Vacca 1994; NGC 4038/4039: Whitmore \& Schweizer 1995; NGC 5018: Hilker
\& Kissler-Patig 1996), the results for NGC 1275 and NGC 5018 are included in
Sects.~2 and 3, the other galaxies do not have their whole globular
cluster system investigated yet.

And indeed, the ``bright'' globular cluster systems 
show all the signs expected after a merger event. The specific frequency (or
efficiency in producing globular clusters) is higher, calling for an
extra process of globular cluster formation that could have been the
merger event. In absolute numbers, bright galaxies host more globular clusters 
as shown by the likely break in the relation in Sect.~3.2. The bright systems
have also a flatter density profile than the light of their host
galaxies. Note however that it is not the galaxy light that steepens, but the 
globular cluster systems
that are rather absolutely flatter, which could mean dynamically hotter after a 
merger event (see evidences in NGC 1399, Grillmair et al.~1994). 
Further the color information listed in column 10 of Table 1 show that 
multi--modal color distributions and color gradients in globular cluster
systems are almost exclusively seen in bright galaxies, while faint
galaxies show globular cluster systems with narrower, single peaked color 
distributions. Finally the brighter galaxies have globular cluster
systems with higher mean metallicities that might be explained by
contributions from newly formed, metal--richer globular clusters
(e.g.~Zepf et al.~1993). Thus Ashman \& Zepf's (1992) predictions seem 
fulfilled for the ``bright'' systems. 

\section{Conclusions}

``Faint'' globular cluster systems do not show signs
of perturbations. They globular clusters spatially follow the galaxy light,
their efficiency in producing globular clusters ($S_N$ values) show
little scatter, and their mean metallicity scales with the galaxy
luminosity. This facts are compatible with a formation of the globular clusters
during the early collapse of the galaxy (Searle \& Zinn 1978, Harris \& Pudritz 
1994).

On the other hand ``bright'' globular cluster systems seem to verify all the
signs expected after the formation in merger events as predicted by 
Ashman \& Zepf (1992). The specific frequency scatters up to very high
values, the mean metallicities show a larger scatter with eventual
multi--modal distributions, the spatial distribution of the globular
clusters is more extended. 

This would support a combination of two formation scenarios for globular 
clusters: every galaxy built up a system during the proto--galactic collapse,
however systems in bright galaxies were then enriched during a merger event 
and still carry its signature.

Translated onto the host galaxies: ``faint'' systems appear in faint,
disky early--type galaxies, while ``bright'' systems are hosted by
bright, boxy ellipticals. This would support similar early
suggestion of Bender (1988) and Nieto (1988) that early--type galaxies can 
be divided into those which formed by a major merger event (involving at least
two galaxies), and those which remained largely unperturbed.

Of course this picture needs a more detailed investigation. A major
merger event might not be the only cause for a dichotomy in early--type
galaxies and many different kind of mergers between very different
galaxies can occur. Globular cluster systems after merger events might
show a large diversity: more or less new globular clusters might have
formed (e.g.~Hilker \& Kissler-Patig 1996), one old system can still
largely dominate, galaxies in the center of cluster still exhibit
extremely high $S_N$ values that could have a additional cause, etc...  
Furthermore, ``faint'' globular cluster systems look very similar to the system 
of the Milky--Way (e.g.~Kissler-Patig et al.~1996a), that might not have
been ``unperturbed'' but accreted some smaller companion galaxies that
could have built up part of the halo (e.g.~Minniti et al.~1996a and
references therein). This immediately calls for transition cases between
mergers and accretion of big companions: is the cut between the two
classes really sharp?

Further investigations especially of the often neglected ``faint'' globular 
cluster systems is needed to give 
further support to the dichotomy in globular cluster systems. Making a
reliable model for the formation of the associated galaxies will probably 
have to wait for
detailed spectroscopic data of whole globular cluster systems. 
But approaching the problem
of formation of early--type galaxies from the globular cluster point of view 
seems promising. 

\acknowledgements

Many thanks to Tom Richtler and Georges Meylan for advising me during my
Thesis. Thanks also to Michael Hilker, Boris Dirsch and Sven Kohle from 
the DFG project Ri 418/5-1 for many fruitful
discussions. Many thanks to Bill Harris for providing me the electronic
version of his globular cluster system compilation, and to Ralf Bender
for communicating me some results of isophotal analysis prior to
publication. Many thanks also to Steve Zepf, the referee, for his
suggestions that helped to improve the paper.
I acknowledge the student fellowship of the European Southern Observatory
during which the frame for this work was set, and the DFG Graduierten
Kolleg ``Das Magellansche System und andere Zwerggalaxien'' for a stipend.

\enddocument